\documentclass[aps,prb,twocolumn,floatfix]{revtex4}

\usepackage{graphicx}
\usepackage{amssymb,amsfonts,amsmath}
\usepackage{epsf}
\usepackage{subfigure}
\usepackage{epstopdf}
\usepackage{mathrsfs}

\font\helvb=cmssbx12

\newcommand{\be}{\begin{equation}}
\newcommand{\ee}{\end{equation}}
\newcommand{\bea}{\begin{eqnarray}}
\newcommand{\eea}{\end{eqnarray}}

\begin{document}

\title{\bf Fluctuation theorems for capacitively coupled electronic currents}

\author{Gregory Bulnes Cuetara}
%\email{gbulnesc@ulb.ac.be}
\author{Massimiliano Esposito}
%\email{mesposit@ulb.ac.be}
\author{Pierre Gaspard}
%\email{gaspard@ulb.ac.be}
\affiliation{Center for Nonlinear Phenomena and Complex Systems,\\
Universit\'e Libre de Bruxelles, Code Postal 231, Campus Plaine,
B-1050 Brussels, Belgium}

\begin{abstract}
The counting statistics of electron transport is theoretically studied in a system with two capacitively coupled parallel transport channels.  Each channel is composed of a quantum dot connected by tunneling to two reservoirs.  The nonequilibrium steady state of the system is controlled by two affinities or thermodynamic forces, each one determined by the two reservoirs of each channel.  The status of a single-current fluctuation theorem is investigated starting from the fundamental two-current fluctuation theorem, which is a consequence of microreversibility.  We show that the single-current fluctuation theorem holds in the limit of a large Coulomb repulsion between the two parallel quantum dots, as well as in the limit of a large current ratio between the parallel channels.  In this latter limit, the symmetry relation of the single-current fluctuation theorem is satisfied with respect to an effective affinity that is much lower than the affinity determined by the reservoirs.  This back-action effect is quantitatively characterized.
\end{abstract}

\maketitle

\section{Introduction}

Away from equilibrium, fluctuating currents flow across small open quantum systems such as quantum dots exchanging electrons with reservoirs.  Advances in nonequilibrium statistical mechanics have shown that the current fluctuations obey symmetry relations following from microreversibility and known as fluctuation theorems.\cite{ECM93,GC95,K98,LS99,AG06,AG07JSP,A09}  They have been proved in different contexts and, especially, for open quantum systems and the full counting statistics of electron transport.\cite{TN05,EHM07,HEM07,EHM09,SU08,AGMT09,SLSB10,CTH11}  In this context, fluctuation theorems relate the probabilities of opposite random values of the currents to the potential differences driving the mean values of the currents.  In electronic circuits, these potential differences play the role of thermodynamic forces also called affinities.\cite{DD36,C85}  Fluctuation theorems hold in nonlinear transport regimes, in particular, for the description of the Coulomb drag effect in capacitively coupled quantum dots.\cite{SLSB10}

Remarkably, modern technology is able to perform the bidirectional counting of single-electron transfers in quantum-dot circuits, allowing the experimental verification of the fluctuation theorem.\cite{FHTH06}   In these experiments, the quantum-dot (QD) circuit is monitored by a parallel circuit made of a quantum point contact (QPC).  Because of electrostatic interactions, the electronic occupancy of the quantum dots modifies the mean value of the QPC current, enabling the measurement of the QD electronic state in real time. The surprise has been that, within experimental error, the bidirectional counting of the QD current obeys the symmetry relation predicted by the fluctuation theorem but with respect to an affinity about one order of magnitude smaller than the potential difference driving the QD circuit.\cite{UGMSFS10}   This discrepancy has revealed the importance of the interaction between the QD and QPC circuits.  Indeed, the QPC current is typically $10^7$-$10^8$ times larger than the QD current in such experiments so that the QPC can act as a quasi-classical detector measuring the quantum state of the QDs.  As a consequence, the whole system composed of the two parallel circuits is quite far from equilibrium and the shot noise in the QPC current has a significant back action onto the small QD current.  In Ref.~\onlinecite{UGMSFS10}, this back action was analyzed in terms of the so-called $P(E)$ theory\cite{IN92} by fitting experimental data to a simple Lorentzian in order to take into account the global effect of the QPC noise onto the QD tunneling rates.  In Ref.~\onlinecite{UGMSFS10bis}, a stochastic model was proposed with extra QPC states besides the QD states.  Both approaches leave open the fundamental understanding of the back action in terms of the microscopic Hamiltonian describing the interaction between the QD and QPC circuits.

\begin{figure}[htbp]
\centerline{\includegraphics[width=8cm]{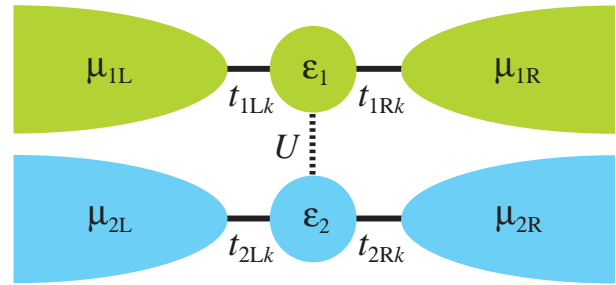}}
\caption{Schematic representation of two quantum dots in parallel.  Each quantum dot is coupled
to two reservoirs of electrons.  Moreover, both quantum dots influence each other by the Coulomb electrostatic interaction.}
\label{fig1}
\end{figure}

In the present paper, we address this issue by considering a system composed of two capacitively coupled parallel transport channels, each containing a single QD in contact with two electron reservoirs.\cite{SLSB10,SKB10,MDZEMHG07}   The two parallel transport channels are only coupled by the electrostatic Coulomb repulsion between the electrons occupying the two QDs so that there is no electron transfer between both channels (see Fig. \ref{fig1}).  The two currents flowing in parallel across this device are driven by two affinities defined by the potential differences on both QDs.  In the following, the circuit No.\,1 plays the role of the QD and the circuit No.\,2 the role of the QPC.  The fluctuations of the two currents obey a bivariate fluctuation theorem, which is the fundamental consequence of microreversibility.  Under general conditions, this two-current fluctuation theorem does not imply the existence of a single-current fluctuation theorem for the QD current monitored by the secondary QPC circuit.  Here, we show that the single-current fluctuation theorem only holds in the limit where the current in the secondary circuit is much larger than the one in the main circuit (or vice versa).  However, the symmetry of the single-current fluctuation theorem does not hold with respect to the potential difference on the main circuit but to an effective affinity which strongly depends on the electrostatic interaction between both circuits.  In this way, our analysis provides an understanding of these features in terms of the basic parameters of the system Hamiltonian and clearly shows that the effective affinity can vary by one order of magnitude or more due to the back action of the QPC onto the QD.

Furthermore, our analysis leads to the evaluation of the entropy production in the electronic device.
The fluctuation theorem has for consequence the non-negativity of the entropy production and is thus compatible with the second law of thermodynamics.  The directionality due to the nonequilibrium driving of the device
is characterized by the probability distributions of the current fluctuations, by the mean values of the currents, and also by the entropy production.  The analysis based on the fluctuation theorem allows us to understand the connections between these complementary and fundamental aspects of such nonequilibrium electronic devices.

The paper is organized as follows.  In Sec. \ref{Model}, the Hamiltonian model is presented and the master equation ruling the occupancies of the QDs is derived for QDs weakly coupled to the reservoirs within the Markovian and secular approximations.  Section \ref{FCS-FT} is devoted to the full counting statistics of the two interacting currents, for which the fundamental fluctuation theorem is established.  Moreover, the connection between the fluctuation theorem and the entropy production of the device is discussed.  In Sec. \ref{Uinfty}, we first consider the limit of a large Coulomb repulsion between the QDs, in which case a single-current fluctuation theorem is obtained but without modification of the effective affinity contrary to the experimental observation.  In Sec. \ref{Limit}, the limit is then considered where the current in one circuit is much larger than the one in the other circuit.  It is in this limit that the single-current fluctuation theorem is obtained with an important modification of the effective affinity with respect to which the symmetry relation of the single-current fluctuation theorem holds.  In Sec. \ref{Numerics}, these effects are numerically demonstrated with the model for parameter values corresponding to typical experimental conditions.   We analyze the dependence of the effective affinity on the parameters of the Hamiltonian model and, especially, on the electrostatic interaction between both circuits.  Conclusions are drawn in Sec. \ref{Conclusions}.

\section{Capacitively coupled parallel transport channels}
\label{Model}

\subsection{The Hamiltonian}

The vehicle of our study is the Hamiltonian model considered in Ref.~\onlinecite{SKB10}.
Each transport channel ($\alpha=1$ or $\alpha=2$) is composed of one quantum dot with a single energy level $\epsilon_\alpha$ for the electron.  
This level is either occupied or empty and the spin degree of freedom is ignored.
Moreover, the quantum dots are capacitively coupled by electrostatic repulsion if both are occupied.  This electrostatic repulsion is taken into account by an Anderson-type term with the parameter $U$.
The parameter $U$ is thus the energy contribution of the Coulomb repulsion when both quantum dots are occupied by an electron.  The system Hamiltonian is therefore given by
\be
H_{\rm S} = \epsilon_1 \, d_1^{\dagger} d_1 +\epsilon_2 \, d_2^{\dagger} d_2 +U d_1^{\dagger} d_1 d_2^{\dagger} d_2
\label{H_S}
\ee
where $d_\alpha$ and $d_\alpha^{\dagger}$ denote the annihilation and creation operators of an electron on the QD labeled by $\alpha=1,2$.  This Hamiltonian is diagonalized in the four-state basis $\{\vert 00\rangle,\vert 10\rangle,\vert 01\rangle,\vert 11\rangle\}$ with the corresponding energy eigenvalues $\{0,\epsilon_1,\epsilon_2,\epsilon_1+\epsilon_2+U\}$.

Each QD is in tunneling contact with two reservoirs on its left- and right-hand sides (see Fig.~\ref{fig1}).
The system has thus four reservoirs $j=1{\rm L},1{\rm R},2{\rm L},2{\rm R}$, which are denoted as $j=\alpha i$
by the label $\alpha=1,2$ of the channel and the side $i={\rm L},{\rm R}$ where the reservoir stands.
The Hamiltonian of all the reservoirs can be expressed as
\be
H_{\rm R} = \sum_j H_j 
\label{H_R}
\ee
in terms of the Hamiltonians of the individual reservoirs, which are defined as
\be
H_j = \sum_k \epsilon_{jk}\, c_{jk}^{\dagger} c_{jk}
\label{H_j}
\ee
where $c_{jk}$ and $c_{jk}^{\dagger}$ denote the annihilation and creation operators of electrons in the corresponding states.  The reservoirs are supposed to be much larger than the system itself so that the eigenvalues $\{\epsilon_{jk}\}$ of each reservoir form a very dense spectrum which is quasi continuous and characterized by a density of states $D_j(\epsilon)=\sum_k \delta(\epsilon-\epsilon_{jk})$.  The operator giving the electron number in the reservoir $j$ is furthermore defined as $N_j=\sum_k c_{jk}^{\dagger} c_{jk}$.

The tunneling Hamiltonian establishing the interaction between the QDs and the reservoirs has the form:
\be
H_{\rm SR} = \sum_{\alpha=1,2}\sum_{i={\rm L},{\rm R}}\sum_k t_{\alpha i k} \, d_{\alpha}^{\dagger} c_{\alpha i k} + {\rm H.\,c.}
\label{H_SR}
\ee
where we have here specified the channels and the reservoirs by writing $j=\alpha i$.
The effect of the electrostatic interaction on the energy barriers between the quantum dots and the reservoirs could be taken into account by including corresponding capacitances, as considered in Ref.~\onlinecite{SLSB10}.  This effect is neglected in the Hamiltonian model of Ref.~\onlinecite{SKB10} that we here consider.

Finally, the total Hamiltonian is defined as the sum:
\be
H=H_{\rm S}+H_{\rm R}+H_{\rm SR}
\label{H}
\ee

We notice that the electron number operators of each transport channel
\be
N_\alpha = d_\alpha^{\dagger} d_\alpha +\sum_{i={\rm L},{\rm R}}\sum_k c_{\alpha i k}^{\dagger} c_{\alpha i k} \qquad \alpha=1,2
\ee
separately commutes with the total Hamiltonian
\be
[ H,N_1 ] = [H, N_2 ] = 0
\label{c=2}
\ee
so that the electron number is conserved on each transport channel and there is no electron exchange between the channels.  In contrast, the number operators of the reservoirs $N_j=N_{\alpha i}$ with $\alpha=1,2$ and $i={\rm L},{\rm R}$ do not commute with the total Hamiltonian unless the tunneling amplitudes are equal to zero.

\subsection{The master equation}

Initially, the reservoirs are in grand-canonical equilibrium states characterized by the chemical potentials $\mu_j$ with $j\in\{ 1{\rm L},1{\rm R},2{\rm L},2{\rm R}\}$ and a uniform temperature $T$.  We denote by $\beta=(k_{\rm B}T)^{-1}$ the inverse temperature with the Boltzmann constant $k_{\rm B}$.  On the other hand, the QDs are in an arbitrary statistical mixture $\rho_{\rm S}(0)$.  Moreover, a measurement is performed at the initial time that determines the numbers $m_1$ and $m_2$ of electrons in the reservoirs $j=1{\rm L}$ and $j=2{\rm L}$.  Consequently, the initial density matrix of the total system has the factorized form:
\be
\rho_{m_1m_2}(0)= \rho_{\rm S}(0)  \prod_{j} \frac{1}{\Xi_j} \, {\rm 
e}^{-\beta(H_j-\mu_{j} N_j)}  \, \delta_{N_{1{\rm L}},m_1}  \, \delta_{N_{2{\rm L}},m_2}
\label{rho_0}
\ee
where $\Xi_j$ denotes the partition function of the grand-canonical ensemble for the reservoir $j$.
The Kronecker symbols $\delta_{N,m}$ take the unit value if $N=m$ and zero otherwise and they thus play the role of projection operators on states with a fixed number of particles.
Thereafter, the density matrix of the total system evolves in time according to the Landau-von Neumann equation
\be
i\,\partial_t \,\rho_{m_1m_2}(t) =[ H,\rho_{m_1m_2}(t)]
\label{LvN_eq}
\ee
in units where $\hbar=1$.  The following normalization condition is satisfied by the initial density matrix (\ref{rho_0}) and preserved by the time evolution (\ref{LvN_eq}):
\be
\sum_{m_1,m_2}{\rm Tr} \,\rho_{m_1m_2}(t)=1
\ee
where $\rm Tr$ denotes the trace over all the degrees of freedom of the total system.

Since we are interested in the occupancies of the QDs and the numbers of electrons transferred between the reservoirs, we focus on the probabilities $p_{\nu_1\nu_2}(n_1,n_2)$ that the QDs are in the quantum states $\{ \vert\nu_1\nu_2\rangle\}$ with the occupancies $\nu_1=0,1$ and $\nu_2=0,1$, while $n_1$ electrons have been transferred from the reservoir $j=1{\rm L}$ to the first QD and $n_2$ electrons from the reservoir $j=2{\rm L}$ to the second QD between the initial time $t=0$ and the time $t$.  These probabilities can be defined in terms of the density matrix of the total system according to
\bea
&& p_{\nu_1\nu_2}(n_1,n_2)= \sum_{m_1,m_2} {\rm Tr}\Big[\rho_{m_1m_2}(t) \,\vert\nu_1\nu_2\rangle\langle \nu_1\nu_2\vert \nonumber\\
&& \qquad \qquad \qquad \qquad  \times \, \delta_{N_{1{\rm L}},m_1-n_1} \, \delta_{N_{2{\rm L}},m_2-n_2}\Big] 
\label{p(n1,n2)}
\eea
which results from a second measurement at time $t$ counting the numbers of transferred electrons.\cite{EHM09}

We suppose that the two quantum dots are weakly coupled to the reservoirs by small enough tunneling amplitudes $\{ t_{jk}\}$ so that we may carry out the Born perturbative approximation on the Landau-von Neumann equation up to second order in the tunneling amplitudes.  We use the secular (or rotating wave) approximation and we take the Markovian approximation.\cite{EHM09,CDG96,SB08,HEM06}  As a consequence, the charging and discharging transition rates of the QDs are respectively given by
\bea
&& a_j = \Gamma_j f_j  \label{aj}\\
&& \bar{a}_j = \bar\Gamma_j \bar{f}_j \label{aUj}\\
&& b_j = \Gamma_j (1-f_j) \label{bj}\\
&& \bar{b}_j = \bar\Gamma_j (1-\bar{f}_j) \label{bUj}
\eea
in terms of the Fermi-Dirac distributions
\bea
&& f_j = \frac{1}{1+{\rm e}^{\beta(\epsilon_j-\mu_j)}} \label{fj}\\
&& \bar{f}_j = \frac{1}{1+{\rm e}^{\beta(\epsilon_j+U-\mu_j)}} \label{fUj}
\eea
where $\epsilon_j=\epsilon_\alpha$ for $j=\alpha i$.
The rate constants are given at the second order of perturbation theory by
\bea
&\Gamma_j &= 2\pi \sum_k \vert t_{jk}\vert^2 \delta(\epsilon_j-\epsilon_{jk}) \nonumber\\
 &&= 2\pi \vert t_j(\epsilon_j)\vert^2 D_j(\epsilon_j)  \\\nonumber\\
&\bar\Gamma_j &= 2\pi \sum_k \vert t_{jk}\vert^2 \delta(\epsilon_j+U-\epsilon_{jk}) \nonumber\\
 &&= 2\pi \vert t_j(\epsilon_j+U)\vert^2 D_j(\epsilon_j+U)
\eea
where the quantities $t_j(\epsilon)$ are the tunneling amplitudes as a function of energy and $D_j(\epsilon)$ the density of states of the reservoir $j$.

The total system is characterized by two sets of time scales:

(1) The correlation times of the reservoirs:  The correlation time of the reservoir $j$ can be estimated as $\tau_{j}^{\rm (C)}\sim \Delta\epsilon_j^{-1}$ in terms of the width $\Delta\epsilon_j$ of the function giving the charging rate $a_j(\epsilon)=2\pi\vert t_j(\epsilon)\vert^2 D_j(\epsilon) f_j(\epsilon)$ versus the energy $\epsilon$.

(2) The relaxation times induced by the electron exchanges with the reservoirs: $\tau_j^{\rm (R)}\sim \Gamma_j^{-1}$.

In consistency with the assumption of weak coupling, we suppose that the correlation times are much shorter than the relaxation times and that the secular approximation is performed by averaging the equation of motion over a time scale $\Delta t$ which is intermediate between both
\be
\tau_{j}^{\rm (C)} \ll \Delta t \ll \tau_{j}^{\rm (R)}
\label{Dt_begin}
\ee
which justifies the use of the Markovian approximation.\cite{CDG96,SB08}  Moreover, the Lamb shifts of the QD energy levels are neglected for simplicity.  

Accordingly, the master equation for the probabilities
\be
{\bf p}(n_1,n_2)=
\left(
\begin{array}{c}
p_{00}(n_1,n_2) \\
p_{10}(n_1,n_2) \\
p_{01}(n_1,n_2) \\
p_{11}(n_1,n_2)
\end{array}
\right)
\ee
takes the form
\begin{widetext}
\be
\partial_t\,{\bf p}(n_1,n_2) = \left(\hat{\mbox{\helvb L}}_1+\hat{\mbox{\helvb L}}_2\right) \cdot {\bf p}(n_1,n_2) 
\label{master}
\ee
with the matricial operators:
\be
\hat{\mbox{\helvb L}}_1
=
\left(
\begin{array}{cccc}
-a_{1{\rm L}}-a_{1{\rm R}} & b_{1{\rm L}}\,\hat{E}_1^+ +b_{1{\rm R}} & 0 & 0 \\
a_{1{\rm L}}\,\hat{E}_1^- +a_{1{\rm R}} & -b_{1{\rm L}}-b_{1{\rm R}} & 0 & 0 \\
0 & 0 & -\bar{a}_{1{\rm L}}-\bar{a}_{1{\rm R}} & \bar{b}_{1{\rm L}}\,\hat{E}_1^+
+\bar{b}_{1{\rm R}}  \\
0 & 0 & \bar{a}_{1{\rm L}}\,\hat{E}_1^- +\bar{a}_{1{\rm R}} & -\bar{b}_{1{\rm L}}-\bar{b}_{1{\rm R}}  \\
\end{array}
\right)
\label{L1}
\ee
and
\be
\hat{\mbox{\helvb L}}_2
=
\left(
\begin{array}{cccc}
-a_{2{\rm L}}-a_{2{\rm R}} & 0 &b_{2{\rm L}}\,\hat{E}_2^+ +b_{2{\rm R}} &  0 \\
0 & -\bar{a}_{2{\rm L}}-\bar{a}_{2{\rm R}} 
& 0 &\bar{b}_{2{\rm L}} \,\hat{E}_2^+ +\bar{b}_{2{\rm R}}\\
a_{2{\rm L}} \,\hat{E}_2^- +a_{2{\rm R}} & 0 &-b_{2{\rm L}}-b_{2{\rm R}} &  0 \\
0 & \bar{a}_{2{\rm L}} \,\hat{E}_2^- +\bar{a}_{2{\rm R}} & 0 &-\bar{b}_{2{\rm L}}-\bar{b}_{2{\rm R}}\\
\end{array}
\right)
\label{L2}
\ee
\end{widetext}
where the step operators
\be
\hat{E}_\alpha^{\pm} \equiv \exp\left( \pm \frac{\partial}{\partial n_\alpha} \right)
\label{E±}
\ee
increase or decrease the numbers $n_\alpha$ of transferred electrons
\be
\hat{E}_\alpha^{\pm} f(n_\alpha) = f(n_\alpha\pm 1)
\ee
when applied on any function $f(n_\alpha)$.\cite{vK81}

We notice that the occupancy probabilities irrespective of the numbers of transferred electrons defined as
\be
P_{\nu_1\nu_2}=\sum_{n_1=-\infty}^{+\infty}\sum_{n_2=-\infty}^{+\infty} p_{\nu_1\nu_2}(n_1,n_2)
\label{Probs}
\ee
obey the master equation obtained by replacing the step operators (\ref{E±}) by unity, $\hat{E}_\alpha^{\pm}=1$,
in the matricial operators (\ref{L1}) and (\ref{L2}).  

\section{The two-current fluctuation theorem and its consequences}
\label{FCS-FT}

\subsection{The cumulant generating function and the affinities}

In order to perform the counting statistics of the electrons transferred from the left reservoirs to the quantum dots, we introduce the cumulant generating function of the currents in terms of the counting parameter $\lambda_\alpha$ of the corresponding transport channel:
\be
Q(\lambda_1,\lambda_2) \equiv \lim_{t\to\infty} - \frac{1}{t} \ln \left\langle \exp(-\lambda_1 n_1 - \lambda_2 n_2 ) \right\rangle_t 
\label{Q}
\ee
where the average
\be
\left\langle X\right\rangle \equiv \sum_{\nu_1,\nu_2,n_1,n_2} p_{\nu_1\nu_2}(n_1,n_2) X
\label{average}
\ee
is taken with respect to the probability distribution, which is the solution of the master equation (\ref{master}) at the time $t$.

We notice that the cumulant generating function (\ref{Q}) is given as the leading eigenvalue of the following eigenvalue problem:
\be
\mbox{\helvb L} \cdot {\bf v} = - Q  \, {\bf v} 
\label{eigenvalue_problem}
\ee
where
\be
\mbox{\helvb L} \equiv {\rm e}^{-\pmb{\lambda}\cdot{\bf n}} \left( \hat{\mbox{\helvb L}}_1+\hat{\mbox{\helvb L}}_2\right){\rm e}^{+\pmb{\lambda}\cdot{\bf n}} = \mbox{\helvb L}_1 + \mbox{\helvb L}_2
\label{L}
\ee
is a four-by-four matrix with real elements that depend on the counting parameters $\pmb{\lambda}$.
Since the functions $\exp(\pmb{\lambda}\cdot{\bf n})$ are the eigenfunctions of the step operators (\ref{E±}),
the matrices $\mbox{\helvb L}_1$ and $\mbox{\helvb L}_2$ are obtained by the following substitutions
\be
\hat{E}_\alpha^{\pm} \to {\rm e}^{\pm\lambda_\alpha}
\ee
in Eqs. (\ref{L1}) and (\ref{L2}), as can be checked by a straightforward calculation.

The four-by-four matrix $\mbox{\helvb L}=\mbox{\helvb L}(\pmb{\lambda})$ obeys the symmetry
\be
\mbox{\helvb M}^{-1} \cdot \mbox{\helvb L}(\pmb{\lambda}) \cdot \mbox{\helvb M}
= \mbox{\helvb L}({\bf A}-\pmb{\lambda})^{\rm T}
\label{symmetry}
\ee
with
\begin{widetext}
\be
\mbox{\helvb M} = 
\left(
\begin{array}{cccc}
1 & 0 & 0 &  0 \\
0 & {\rm e}^{-\beta(\epsilon_1-\mu_{1{\rm R}})} & 0 & 0 \\
0 & 0 & {\rm e}^{-\beta(\epsilon_2-\mu_{2{\rm R}})} & 0 \\
0 & 0 & 0 & {\rm e}^{-\beta(\epsilon_1+\epsilon_2+U-\mu_{1{\rm R}}-\mu_{2{\rm R}})} \\
\end{array}
\right)
\ee
\end{widetext}
and the affinities ${\bf A}=(A_1,A_2)$ defined by
\bea
A_1 = \ln\frac{a_{1{\rm L}}b_{1{\rm R}}}{b_{1{\rm L}}a_{1{\rm R}}} 
= \ln\frac{\bar{a}_{1{\rm L}}\bar{b}_{1{\rm R}}}{\bar{b}_{1{\rm L}}\bar{a}_{1{\rm R}}} 
= \beta \left( \mu_{1{\rm L}} -\mu_{1{\rm R}} \right) \label{A1}\\
A_2 = \ln\frac{a_{2{\rm L}}b_{2{\rm R}}}{b_{2{\rm L}}a_{2{\rm R}}} 
= \ln\frac{\bar{a}_{2{\rm L}}\bar{b}_{2{\rm R}}}{\bar{b}_{2{\rm L}}\bar{a}_{2{\rm R}}} 
= \beta \left( \mu_{2{\rm L}} -\mu_{2{\rm R}} \right) \label{A2} 
\eea
We notice that these affinities can also be obtained by using Schnakenberg graph analysis.\cite{AG07JSP,S76}
These quantities are the two independent thermodynamic forces able to drive the system away from equilibrium.
The fact that there exists only two independent affinities although the system contains four reservoirs is due to the existence of the two constants of motion (\ref{c=2}) given by the particle numbers in the two transport channels.

If the system was fully connected, only the total particle number would be a constant of motion and there would exist three independent affinities.  More generally, a system composed of $r$ reservoirs and partitioned into $c$ disconnected but interacting transport channels has $c$ constant particle numbers and can be driven away from equilibrium by $r-c$ independent affinities.  Here, $r=4$ and $c=2$ so that there is only $r-c=2$ independent affinities.

As aforementioned, the cumulant generating function is given by the leading eigenvalue of Eq. (\ref{eigenvalue_problem}), i.e., by the smallest root of the quartic characteristic polynomial:
\be
\det\left(\mbox{\helvb L} + Q \, \mbox{\helvb 1}\right)=0
\label{det}
\ee
of the four-by-four matrix (\ref{L}).
Therefore, the symmetry (\ref{symmetry}) implies that the cumulant generating function also obeys this symmetry.\cite{K98,A09}  In this way, the fundamental result is proved that the cumulant generating function satisfies the {\it fluctuation theorem}:
\be
Q(\pmb{\lambda})=Q({\bf A}-\pmb{\lambda})
\ee
or
\be
Q(\lambda_1,\lambda_2)=Q(A_1-\lambda_1,A_2-\lambda_2)
\label{FT}
\ee
in terms of the affinities ${\bf A}=(A_1,A_2)$ given by Eqs. (\ref{A1}) and (\ref{A2}).

An alternative expression of this fluctuation theorem is that the probability
\be
p(n_1,n_2)=\sum_{\nu_1,\nu_2} p_{\nu_1\nu_2}(n_1,n_2)
\ee
for the transfer of $n_1$ particles in the circuit No.\,1 and $n_2$ particles in the circuit No.\,2 during the time interval $t$ obeys
\be
\frac{p(n_1,n_2)}{p(-n_1,-n_2)} \simeq \exp(A_1n_1+A_2n_2) \qquad \mbox{for} \quad t\to+\infty
\label{FT-p}
\ee
Indeed, this expression implies Eq. (\ref{FT}) using the definition (\ref{Q}) of the cumulant generating function with the average (\ref{average}).\cite{AG07JSP,EHM09} 

In general, this two-current fluctuation theorem does not imply any single-current fluctuation theorem unless specific conditions are satisfied either by construction,\cite{AG07} or in some particular limit, which is here the case as shown in the following sections.

\subsection{The average currents and the response coefficients}

The average values of the particle currents are given in terms of the generating function according to
\be
J_{\alpha} = \left.\frac{\partial Q}{\partial\lambda_{\alpha}}\right\vert_{\pmb{\lambda}=0} 
\label{J_a}
\ee
for $\alpha=1,2$. As shown in Appendix \ref{AppA}, these currents can be expressed in terms of the probabilities (\ref{Probs}) of the four QD states in the nonequilibrium steady state corresponding to the affinities (\ref{A1}) and (\ref{A2}) according to:
\bea
&& J_1 = a_{\rm 1L} P_{00} - b_{\rm 1L} P_{10} + \bar{a}_{\rm 1L} P_{01} - \bar{b}_{\rm 1L} P_{11} \label{J1} \\
&& J_2 = a_{\rm 2L} P_{00} - b_{\rm 2L} P_{01} + \bar{a}_{\rm 2L} P_{10} - \bar{b}_{\rm 2L} P_{11} \label{J2}
\eea
These currents are nonlinear functions of the affinities, which can be expanded in powers of the affinities in order to identify the linear and nonlinear response coefficients:
\bea
J_\alpha &=& J_{\alpha}(A_1,A_2) \nonumber\\
&=& \sum_\beta L_{\alpha,\beta}A_\beta
+\frac{1}{2}\sum_{\beta,\gamma} M_{\alpha,\beta\gamma}A_\beta A_\gamma \nonumber\\
&&+\frac{1}{6}\sum_{\beta,\gamma,\delta} N_{\alpha,\beta\gamma\delta}A_\beta A_\gamma A_\delta + \cdots
\eea
As a consequence of the fluctuation theorem (\ref{FT}), the linear response coefficients $L_{\alpha,\beta}$ are given in terms of the second derivatives of the generating  function with respect to the counting parameters and they thus obey the Onsager reciprocity relations:
\be
L_{\alpha,\beta}=L_{\beta,\alpha}=- \frac{1}{2} \frac{\partial^2Q}{\partial\lambda_\alpha\partial\lambda_\beta}\Big\vert_{\pmb{\lambda}=0,{\bf A}=0}
\label{L12}
\ee
Similar relationships have been established for the nonlinear response coefficients.\cite{AG07JSM}

The average currents as well as the linear response coefficients can be calculated in terms of the characteristic determinant (\ref{det}) of the matrix (\ref{L}) as shown in Appendix \ref{AppA}.  By using Eq. (\ref{L12_det}), the Onsager coefficient turns out to be proportional to
\be
L_{1,2}\propto \left(\Gamma_{\rm 1L} \bar{\Gamma}_{\rm 1R}-\bar{\Gamma}_{\rm 1L} \Gamma_{\rm 1R}\right)
\left(\Gamma_{\rm 2L} \bar{\Gamma}_{\rm 2R}-\bar{\Gamma}_{\rm 2L} \Gamma_{\rm 2R}\right)
\label{L12_prop}
\ee
In general, the Onsager coefficient is thus non vanishing and there is a phenomenon of Coulomb drag according to which a current may be induced in a circuit at equilibrium if the other circuit is out of equilibrium, as shown in Ref.~\onlinecite{SLSB10}.  

However, the Onsager coefficient vanishes under the condition that the rate constants of one circuit do not depend on the Coulomb repulsion parameter $U$.  In this particular case, there is no Coulomb drag because
\be
J_1(0,A_2)=0 \qquad \mbox{and} \qquad J_2(A_1,0)=0 \qquad \mbox{if}\quad \Gamma_j = \bar{\Gamma}_j
\label{J_a=0}
\ee
This property is also proved in the Appendix \ref{AppA}.  Equation (\ref{J_a=0}) implies the vanishing of the Onsager coefficient as well as the nonlinear response coefficients allowing the coupling of one current to the affinity of the other circuit:
\bea
L_{1,2}=M_{1,22}=N_{1,222}=\cdots =0  && \mbox{and} \nonumber\\
L_{2,1}=M_{2,11}=N_{2,111}=\cdots =0  && \mbox{if} \quad \Gamma_j = \bar{\Gamma}_j
\label{LMN}
\eea
Nevertheless, these coefficients do not vanish in general.

\subsection{The entropy production and the energy dissipation}

A further consequence of the fluctuation theorem (\ref{FT}) is that the average currents (\ref{J_a}) obeys the second law of thermodynamics according to which the entropy production is always non-negative:
\be
\frac{1}{k_{\rm B}} \frac{d_{\rm i}S}{dt} = A_1 J_1 + A_2 J_2 \geq 0 
\ee
where $k_{\rm B}$ is Boltzmann's constant.\cite{AG07JSP,EHM09}

The power dissipated in each circuit is defined as the product of the voltage $V_\alpha$ by the electric current $I_\alpha=eJ_\alpha$ where $e$ is the electric charge of the particle: $\Pi_\alpha=V_\alpha I_\alpha$, with $\alpha=1,2$.  Since the affinities are related to the voltages by
\be
A_\alpha = \frac{eV_\alpha}{k_{\rm B} T}
\ee
we have that the dissipated power in the circuit $\alpha$ is given by
\be
\Pi_\alpha = k_{\rm B} T \, A_\alpha J_\alpha 
\ee
and the entropy production is thus proportional to the total dissipated power:
\be
\frac{d_{\rm i}S}{dt} = \frac{1}{T} \left( \Pi_1 + \Pi_2 \right) \geq 0 
\ee
Therefore, the entropy production of the system characterizes the energy dissipation of the quantum measurement performed on one QD by the current flowing in the other circuit playing the role of the QPC in the experiments of Refs.~\onlinecite{FHTH06,GLSSISEDG06}.  We shall evaluate these quantities under such specific conditions in the following sections.

\section{The large Coulomb repulsion limit}
\label{Uinfty}

In the limit where the Coulomb repulsion between both QDs is large, the coupling parameter $U$ takes large values so that the charging rates of a second electron on the two QDs vanish:
\be
\bar{a}_j=0 \qquad \mbox{for} \quad j=1{\rm L}, 1{\rm R}, 2{\rm L}, 2{\rm R} \qquad \mbox{if} \quad U=\infty
\label{bara}
\ee
As a consequence, the probabilities (\ref{p(n1,n2)}) and (\ref{Probs}) 
that the system is in the fourth state $\vert 11\rangle$ also vanish:
\be
p_{11}(n_1,n_2)=0 \qquad \mbox{and} \qquad  P_{11}=0 \qquad \mbox{if} \quad U=\infty
\label{P11=0}
\ee

In this limit, the occupancy of one QD is stopping the current in the other QD.  For instance, the average current in the secondary circuit (\ref{J2}) has two contributions depending on the occupancy of the QD No.\,1:
\be
J_2 = \left. J_2\right\vert_{\nu_1=0} + \left. J_2\right\vert_{\nu_1=1} 
\ee
However, the contribution when the QD No.\,1 is occupied is vanishing
\be
\left. J_2\right\vert_{\nu_1=1} = \bar{a}_{\rm 2L} P_{10} - \bar{b}_{\rm 2L} P_{11} = 0 \qquad\mbox{if} \quad U=\infty
\ee
since $\bar{a}_{\rm 2L}=0$ according to Eq. (\ref{bara}) and $P_{11} = 0$ because of Eq. (\ref{P11=0}).
Therefore, the secondary circuit has a non-vanishing current only when the QD No.\,1 is empty and vice versa.

The cumulant generating function can be obtained by considering the three-by-three matrix obtained by removing the fourth row and column from the matrix (\ref{L}).  In this case, the characteristic determinant (\ref{det}) depends on the counting parameters only in the following combinations:
\bea
&& a_{\rm 1R} b_{\rm 1L} {\rm e}^{\lambda_1} + a_{\rm 1L} b_{\rm 1R} {\rm e}^{-\lambda_1} \\
&& a_{\rm 2R} b_{\rm 2L} {\rm e}^{\lambda_2} + a_{\rm 2L} b_{\rm 2R} {\rm e}^{-\lambda_2}
\eea
which remain invariant under the independent substitutions $\lambda_1 \to A_1-\lambda_1$ and/or $\lambda_2 \to A_2-\lambda_2$ with the affinities (\ref{A1}) and (\ref{A2}).  Consequently, we obtain the symmetry relations:
\bea
&& Q(\lambda_1,\lambda_2)=Q(A_1-\lambda_1,\lambda_2)\nonumber\\
&&=Q(\lambda_1,A_2-\lambda_2)=Q(A_1-\lambda_1,A_2-\lambda_2)
\eea
if $U=\infty$, which implies the single-current fluctuation theorem:
\be
Q(\lambda_1,0)=Q(A_1-\lambda_1,0)
\qquad \mbox{if} \quad U=\infty
\ee
but with the unmodified affinity (\ref{A1}). Therefore, this limit cannot explain the modification of the affinity observed in the experiments reported in Ref.~\onlinecite{FHTH06}.

\section{The large current ratio limit}
\label{Limit}

In counting statistics experiments.\cite{FHTH06,GLSSISEDG06}, the QPC which is used to observe the occupancy of the QD carries a current which is typically much larger than the current in the QD by a huge factor $10^7$-$10^8$. If the QPC is taken as the circuit No.\,2 in the present model, the rate constants of that circuit are much larger than the ones of the circuit No.\,1:
\be
\Gamma_{\rm 1L}, \Gamma_{\rm 1R}, \bar{\Gamma}_{\rm 1L}, \bar{\Gamma}_{\rm 1R} \ll \Gamma_{\rm 2L}, \Gamma_{\rm 2R}, \bar{\Gamma}_{\rm 2L}, \bar{\Gamma}_{\rm 2R} 
\label{conditions}
\ee
Under such circumstances, the relaxation times $\tau_{1i}^{\rm (R)}\sim \Gamma_{1i}^{-1}$ of the circuit No.\,1 are much longer than the relaxation times $\tau_{2i}^{\rm (R)}\sim \Gamma_{2i}^{-1}$ of the circuit No.\,2 and the monitoring of the slow circuit by the fast one is performed over a time scale $\Delta t$ such that
\be
\tau_{2i}^{\rm (R)} \ll \Delta t \ll \tau_{1i}^{\rm (R)}
\label{Dt_real}
\ee
instead of the time scale (\ref{Dt_begin}).

Our aim is here to obtain the cumulant generating function for the counting statistics in the sole circuit No.\,1 without measuring the current in the fast circuit No.\,2, as it is the case in Refs.~\onlinecite{FHTH06,GLSSISEDG06}. 
This amounts to consider the two-current generating function (\ref{Q}) for $\lambda_2=0$.
Accordingly, we focus on the time evolution of the probabilities defined by
\be
p_{\nu_1}(n_1)=\sum_{\nu_2=0,1}\sum_{n_2=-\infty}^{+\infty} p_{\nu_1\nu_2}(n_1,n_2)
\label{p_nu1_n1}
\ee

Since the electron transfers in the circuit No.\,2 are much faster than in the circuit No.\,1, the circuit No.\,2 can be supposed to be in a stationary state during the whole period when the circuit No.\,1 is in a given state.
Such stationary states conditional to the state $\nu_1$ of the QD No.\,1 are obtained by finding the zero eigenvectors of the transition matrix (\ref{L2}) with $\hat{E}_2^{\pm}=1$.  The conditional probabilities $P_{\nu_2\vert\nu_1}$ that the QD No.\,2 has the occupancy $\nu_2$ provided that the QD No.\,1 is in the state $\nu_1$ are given by
\bea
&& P_{0\vert 0} = \frac{b_2}{a_2+b_2} \label{P00}\\
&& P_{1\vert 0} = \frac{a_2}{a_2+b_2} \label{P10}\\
&& P_{0\vert 1} = \frac{\bar{b}_2}{\bar{a}_2+\bar{b}_2} \label{P01}\\
&& P_{1\vert 1} = \frac{\bar{a}_2}{\bar{a}_2+\bar{b}_2} \label{P11}
\eea
with
\bea
&& a_2 = a_{2{\rm L}} + a_{2{\rm R}} \\
&& b_2 = b_{2{\rm L}} + b_{2{\rm R}} \\
&& \bar{a}_2 = \bar{a}_{2{\rm L}} + \bar{a}_{2{\rm R}} \\
&& \bar{b}_2 = \bar{b}_{2{\rm L}} + \bar{b}_{2{\rm R}}
\eea

Under the conditions (\ref{conditions}), the probability that the system is in the state $\vert\nu_1\nu_2\rangle$ and that $n_1$ electrons have been transferred in the circuit No.\,1 factorizes into the probability (\ref{p_nu1_n1}) and the probability of the occupancy $\nu_2$ of the QD No.\,2 conditioned to the occupancy $\nu_1$:
\be
p_{\nu_1\nu_2}(n_1)=p_{\nu_1}(n_1) P_{\nu_2\vert\nu_1}
\ee

Substituting these relations into the master equation (\ref{master}) and summing over $n_2$ and $\nu_2$, we get  the master equations for the probabilities $p_{\nu_1}(n_1)$ as follows:
\bea
\partial_t \, p_0(n_1) &=& -\Big(a_{\rm L}+a_{\rm R}\Big)\, p_0(n_1) + \Big(b_{\rm L}\hat{E}_1^+ +b_{\rm R}\Big)\, p_1(n_1) \nonumber\\ \label{p0t} \\
\partial_t \, p_1(n_1) &=&  \Big(a_{\rm L}\hat{E}_1^- +a_{\rm R}\Big)\, p_0(n_1) - \Big(b_{\rm L}+b_{\rm R}\Big)\, p_1(n_1) \nonumber\\\label{p1t}
\eea
where
\bea
&& a_{\rm L} =a_{1{\rm L}} P_{0\vert 0}+ \bar{a}_{1{\rm L}} P_{1\vert 0} \label{aL} \\
&& a_{\rm R} =a_{1{\rm R}} P_{0\vert 0} + \bar{a}_{1{\rm R}} P_{1\vert 0} \label{aR}\\
&& b_{\rm L} =b_{1{\rm L}} P_{0\vert 1} + \bar{b}_{1{\rm L}} P_{1\vert 1} \label{bL}\\
&& b_{\rm R} =b_{1{\rm R}} P_{0\vert 1} + \bar{b}_{1{\rm R}} P_{1\vert 1} \label{bR}
\eea
are the charging and discharging rates of the first quantum dot
averaged over the conditional stationary probabilities of the second quantum dot.  The master equations (\ref{p0t})-(\ref{p1t}) rule the process in the slow circuit No.\,1 as monitored by the fast circuit No.\,2 over the time scale (\ref{Dt_real}).

Taking a solution of the form $p_{\nu_1}(n_1)\sim \exp(\lambda_1n_1-Q t)$ for Eqs. (\ref{p0t})-(\ref{p1t}), 
the cumulant generating function (\ref{Q}) with $\lambda_2=0$ has thus for approximation the leading eigenvalue of the matrix
\be
\tilde{\mbox{\helvb L}} =
\left(
\begin{array}{cc}
-a_{\rm L}-a_{\rm R} &  b_{\rm L}{\rm e}^{+\lambda_1}+b_{\rm R}  \\
a_{\rm L}{\rm e}^{-\lambda_1}+a_{\rm R} & -b_{\rm L}-b_{\rm R} \\
\end{array}
\right)
\label{tilde_L_0}
\ee
which is given by
\begin{widetext}
\be
Q(\lambda_1,0)\simeq  \frac{1}{2} \left[ 
a_{\rm L}+a_{\rm R} +b_{\rm L}+b_{\rm R} -
\sqrt{\left(a_{\rm L}+a_{\rm R} -b_{\rm L}-b_{\rm R}\right)^2
+ 4 \left( a_{\rm L}{\rm e}^{-\lambda_1}+a_{\rm R}\right)\left(b_{\rm L}{\rm e}^{+\lambda_1}+b_{\rm R}\right)}\right]
\label{tilde_Q_0}
\ee
\end{widetext}
in the limit (\ref{conditions}) where the current in the second quantum dot is much larger than in the first one.
In this limit, the generating function (\ref{tilde_Q_0}) obeys the {\it single-current fluctuation theorem}:
\be
Q(\lambda_1,0) = Q(\tilde A_1- \lambda_1,0)
\label{FT1}
\ee
with the effective affinity for the first quantum dot obtained as
\be
\tilde A_1 \equiv \ln \frac{a_{\rm L} b_{\rm R}}{a_{\rm R} b_{\rm L}}
\label{eff_A_1}
\ee
in terms of the averaged rates (\ref{aL})-(\ref{bR}).  This constitutes the main result of the present paper.

We notice that similar results hold in the other limit where the circuit No.\,1 is much faster than the circuit No.\,2 because both circuits have the same structure and are symmetrically coupled together through the Coulomb repulsion of parameter $U$ in Eq. (\ref{H_S}).

The result (\ref{FT1}) shows that the generating function of the counting statistics in the slow QD No.\,1 has the symmetry of a single-current fluctuation theorem under the experimental conditions (\ref{conditions}) but with respect to the effective affinity (\ref{eff_A_1}).  This latter may differ by orders of magnitude with respect to the affinity (\ref{A1}) driving the circuit out of equilibrium.  The reason for this modification is the back action of the other circuit to which the QD is capacitively coupled.  Indeed, the charging and discharging rates of the QD No.\,1 are averaged over the two possible states of the QD No.\,2 according to Eqs. (\ref{aL})-(\ref{bR}) so that their effective values are modified by the back action of the circuit No.\,2.  This modification of the transition rates is reminiscent of the influence of environmental noises as described by the $P(E)$ theory.\cite{IN92}

In the following section, the importance of the back action is numerically demonstrated by considering specific conditions.

\section{Numerical results}
\label{Numerics}

\subsection{Parameter values}

In typical counting statistics experiment,\cite{FHTH06,UGMSFS10} the affinities take quite large values because the voltages are large with respect to the temperature.  For instance, the voltages $V_{\rm QD}=300\; \mu$V, $V_{\rm QPC}=800\; \mu$V, and the electronic temperature $T=130$ mK are reported in Ref.~\onlinecite{FHTH06}.  In our analogy, we choose the affinities as follows:
\bea
&& A_1 = A_{\rm QD}=\frac{eV_{\rm QD}}{k_{\rm B}T} = 25 \label{A1_act}\\
&& A_2 = A_{\rm QPC}=\frac{eV_{\rm QPC}}{k_{\rm B}T}= 70 \label{A2_act}
\eea
Since the QPC current is reduced by about 10\% if the QD is occupied, the parameter $U$ of the Coulomb repulsion between both QDs can be taken as
\be
\beta U = 32.8
\label{U}
\ee
Moreover, the QPC current is about $10^7$-$10^8$ larger than the QD current.

As aforementioned, the role of the QPC is played by the circuit No.\,2 and the QD by the one of the circuit No.\, 1 in our model.  The energy level of the second quantum dot is supposed to be in the middle between the reservoirs electrochemical potentials and the couplings to the reservoirs are chosen symmetric and independent of the energy.  Under such assumptions, possible parameter values are given by 
\bea
&& \beta\mu_{1{\rm L}} = 25 \label{m1L}\\
&& \beta\mu_{1{\rm R}} = 0 \label{m1R}\\
&& \Gamma_{1{\rm L}}= \Gamma_{1{\rm R}} =\bar{\Gamma}_{1{\rm L}}= \bar{\Gamma}_{1{\rm R}} = 1 \label{G1}\\
&& \beta\mu_{2{\rm L}} = 70 \label{m2L}\\
&& \beta\mu_{2{\rm R}} = 0 \label{m2R}\\
&& \Gamma_{2{\rm L}}= \Gamma_{2{\rm R}} = \bar{\Gamma}_{2{\rm L}}= \bar{\Gamma}_{2{\rm R}} = 10^8 \label{G2}\\
&& \beta\epsilon_2 = 35 \label{e2}
\eea
while the level of the QD No.\,1 has the energy $\epsilon_1$, which may take different values in the following numerical calculations.  The correlation times of the reservoirs are supposed to be short enough for the conditions (\ref{Dt_begin}) to be satisfied in consistency with the perturbative approximation.

\subsection{Stochastic simulations}

The random time evolution of the system can be generated by simulating the stochastic jump process of the master equation (\ref{master}) with Gillespie's algorithm.\cite{G76,G77}  Four possible transitions may occur from each of the four states.  The transition rates are given by Eqs. (\ref{aj})-(\ref{bUj}) with the Fermi-Dirac distributions (\ref{fj})-(\ref{fUj}) and the rate constants (\ref{G1})-(\ref{G2}).  

\begin{figure}[h]
\centerline{\includegraphics[width=9cm]{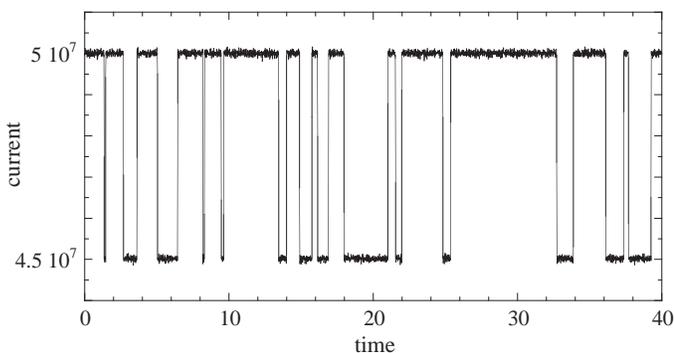}}
\caption{Simulation with Gillespie's algorithm of the QPC current in circuit No. 2 measuring the QD occupancy.  The parameter values are given by Eqs. (\ref{U})-(\ref{e2}) and $\beta\epsilon_1=0$.  The effective affinity of the circuit No. 1 is $\tilde{A}_1=1.17$.  The mean value of the QD current is $J_1\simeq 0.17$ electrons per unit time.  The mean value of the QPC current is $J_2\simeq 4.8\times10^7$ electrons per unit time.  The QD is empty (resp. occupied) when the QPC current takes the value $5\times 10^7$ (resp. $4.5\times 10^7$).}
\label{fig2}
\end{figure}

Figure \ref{fig2} depicts the current in the circuit No.\,2 averaged over a time interval $\Delta t = 0.01$, which is shorter than the typical dwell time of the QD No.\,1, as required by Eq. (\ref{Dt_real}).  We see that the current is reduced by about 10\% when the QD No.\,1 is occupied, which is in agreement with the choice for the parameter (\ref{U}).  The ratio between the mean values of the currents is here given by $J_2/J_1=2.8\times 10^8$, while the ratio of the dissipated powers takes the value $\Pi_2/\Pi_1=(A_2J_2)/(A_1J_1)=7.9\times 10^8$.  Such very large ratios are required in order for the secondary current to distinguish between the two states of the QD in the primary circuit.  Simulations show that the fluctuations of the secondary current would be larger for smaller values of the current ratio.  Thanks to the large ratio, the instantaneous occupancy in the circuit No.\,1 can be monitored by the current in the circuit No.\,2 over the time scale (\ref{Dt_real}), which is longer than the time scale of the fast circuit No.\,2 but shorter than the one of the circuit No.\,1.

\subsection{The cumulant generating function and its properties}

The cumulant generating function of the current in the circuit No.\,1 is calculated by the leading root of the characteristic polynomial (\ref{det}) of the four-by-four matrix (\ref{L}) with $\lambda_2=0$.

\begin{figure}[htbp]
\centerline{\includegraphics[width=8.5cm]{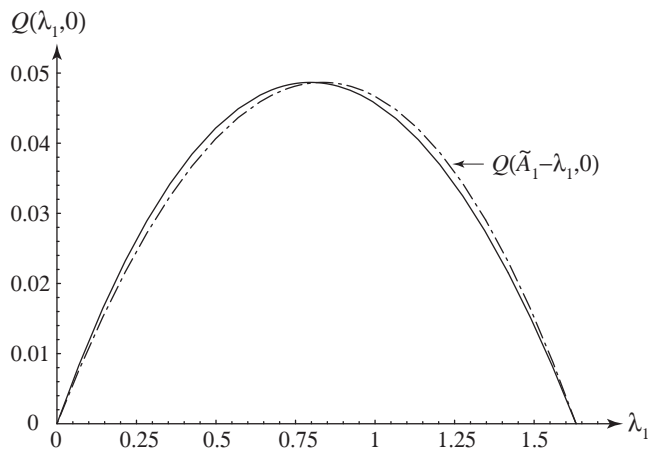}}
\caption{The cumulant generating function versus the counting parameter $\lambda_1$ at $\lambda_2=0$ and the symmetric function with respect to the effective affinity $\tilde{A}_1=1.6319$ (dotted-dashed line) for the parameter values $\beta U=30$, $\beta\epsilon_1=0$, $\beta\epsilon_2=35$, $\beta\mu_{1{\rm L}}=25$, $\beta\mu_{1{\rm R}}=0$, $\beta\mu_{2{\rm L}}=70$, $\beta\mu_{2{\rm R}}=0$, $\Gamma_{1{\rm L}}=\Gamma_{1{\rm R}}=\bar{\Gamma}_{1{\rm L}}= \bar{\Gamma}_{1{\rm R}} =1$, $\Gamma_{2{\rm L}}=\Gamma_{2{\rm R}}= \bar{\Gamma}_{2{\rm L}}= \bar{\Gamma}_{2{\rm R}}=1$.}
\label{fig3}
\end{figure}

The lack of symmetry of the single-current generating function $Q(\lambda_1,0)$ is manifest if the rate constants of both circuits are of the same order of magnitude.  The generating function and its symmetric with respect to the effective affinity is depicted in Fig. \ref{fig3} for $\Gamma_{2s}/\Gamma_{1s}=1$ (with $s={\rm L, R}$) and $\beta U=30$.  Here, the effective affinity is taken as the non-trivial root of the generating function such that $Q(\tilde{A}_1,0)=0$.  We clearly see that the generating function is not symmetric with respect to the effective affinity $Q(\lambda_1,0)\neq Q(\tilde{A}_1-\lambda_1,0)$ so that the single-current fluctuation theorem does not hold in general although the two-current fluctuation theorem always does.  Furthermore, we notice that the effective affinity $\tilde{A}_1=1.6319$ is much smaller than the affinity determined by the reservoirs: $A_1=\beta(\mu_{\rm 1L}-\mu_{\rm 1R})=25$.

\begin{figure}[htbp]
\centerline{\includegraphics[width=8.5cm]{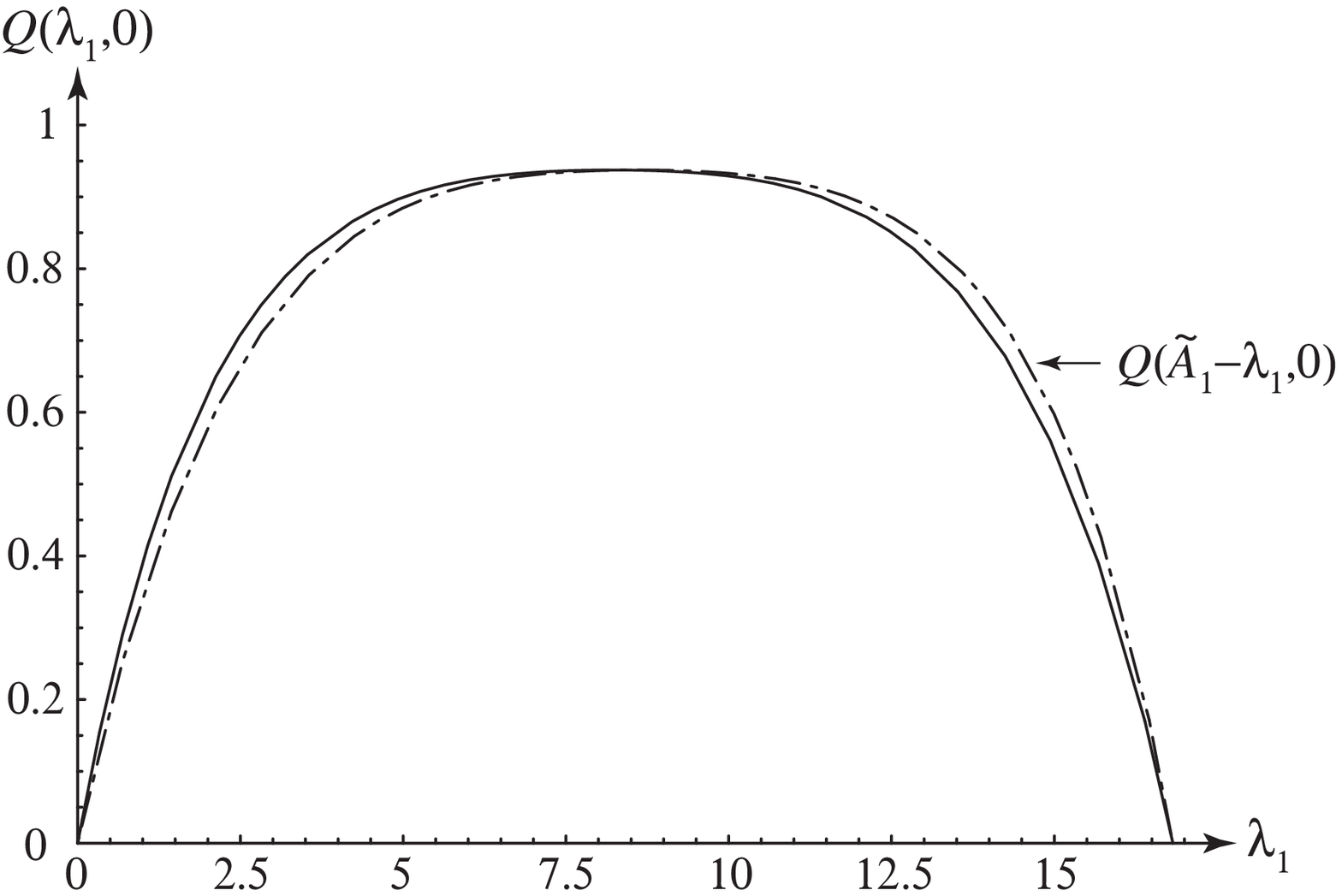}}
\caption{The cumulant generating function versus the counting parameter $\lambda_1$ at $\lambda_2=0$ and the symmetric function with respect to the effective affinity $\tilde{A}_1=16.8356$ (dotted-dashed line) for the parameter values $\beta U=10$, $\beta\epsilon_1=10$, $\beta\epsilon_2=35$, $\beta\mu_{1{\rm L}}=25$, $\beta\mu_{1{\rm R}}=0$, $\beta\mu_{2{\rm L}}=70$, $\beta\mu_{2{\rm R}}=0$, $\Gamma_{1{\rm L}}=\Gamma_{1{\rm R}}=\bar{\Gamma}_{1{\rm L}}= \bar{\Gamma}_{1{\rm R}} =1$, $\Gamma_{2{\rm L}}=\Gamma_{2{\rm R}}= \bar{\Gamma}_{2{\rm L}}= \bar{\Gamma}_{2{\rm R}}=2$.}
\label{fig4}
\end{figure}

In Fig. \ref{fig4}, the single-current generating function is depicted for the smaller value of the Coulomb repulsion $\beta U=10$ and $\Gamma_{2s}/\Gamma_{1s}=2$.  Here, the effective affinity takes a larger value, but again the asymmetry of the generating function is still manifest.  The shape of the generating function now deviates from the parabolic shape seen in Fig. \ref{fig3} as its maximum approaches the unity value.

Although the ratio of the rate constants is of order unity in both Figs. \ref{fig3} and \ref{fig4}, the difference between the generating function and its symmetric is smaller than 5\% and could remain unobservable if the counting statistics was not precise enough.

\begin{figure}[htbp]
\centerline{\includegraphics[width=8cm]{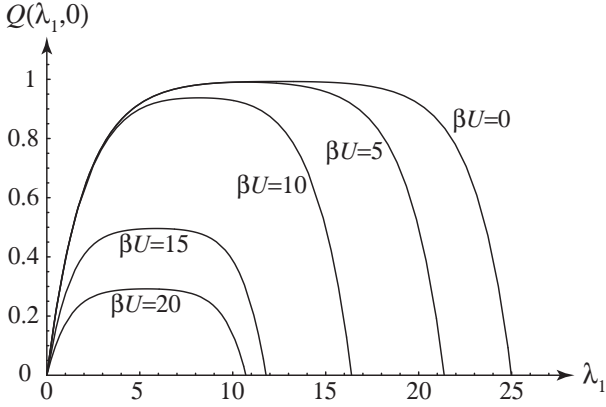}}
\caption{The cumulant generating function versus the counting parameter $\lambda_1$ at $\lambda_2=0$ for different values of the electrostatic coupling parameter $\beta U$. The other parameters take the values $\beta\epsilon_1=10$, $\beta\epsilon_2=35$, $\beta\mu_{1{\rm L}}=25$, $\beta\mu_{1{\rm R}}=0$, $\beta\mu_{2{\rm L}}=70$, $\beta\mu_{2{\rm R}}=0$, $\Gamma_{1{\rm L}}=\Gamma_{1{\rm R}}=\bar{\Gamma}_{1{\rm L}}= \bar{\Gamma}_{1{\rm R}} =1$, $\Gamma_{2{\rm L}}=\Gamma_{2{\rm R}}=100$.}
\label{fig5}
\end{figure}

Figure \ref{fig5} shows the deformation of the generating function $Q(\lambda_1,0)$ as the electrostatic coupling parameter $U$ varies from zero to $\beta U=20$ for $\Gamma_{2s}/\Gamma_{1s}=100$.  In the absence of electrostatic coupling, the single-current fluctuation theorem holds in the circuit No.\,1 since it is decoupled from the rest of the system.  In this case, the affinity takes the value $A_1=25$ determined by the two reservoirs of this circuit, as seen in Fig. \ref{fig5}.  However, the non-trivial root $\tilde{A}_1$ of the generating function decreases as the Coulomb repulsion $U$ increases, showing the back-action effect of the secondary circuit due to the capacitive coupling.  In the same progression, the maximum of the generating function is also reduced.

For the ratio of rate constants taken in Fig. \ref{fig5}, the generating function is already practically indistinguishable from its symmetric $Q(\tilde{A}_1-\lambda_1,0)$ so that the single-current fluctuation theorem is already effective and the considerations of Sec. \ref{Limit} apply.  In particular, the effective affinity is now very well approximated by Eq. (\ref{eff_A_1}).  

\subsection{The large current ratio limit and the effective affinity}

\begin{figure}[h]
\centerline{\includegraphics[width=9cm]{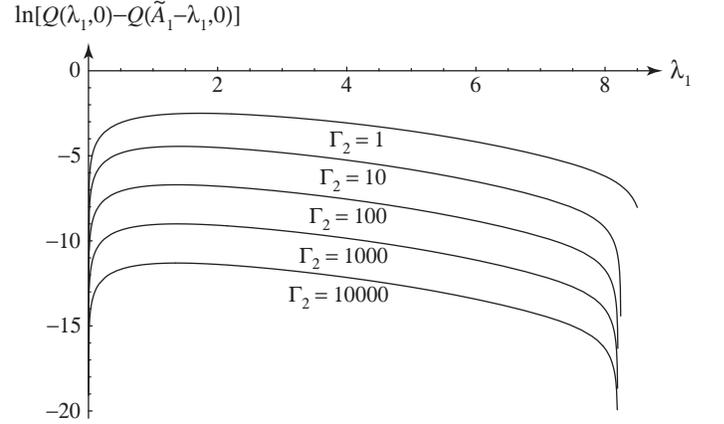}}
\caption{The difference between the cumulant generating function and its symmetric with respect to the effective affinity $\tilde{A}_1$ versus the counting parameter $\lambda_1$ at $\lambda_2=0$ for the parameter values $\beta U=10$, $\beta\epsilon_1=10$, $\beta\epsilon_2=35$, $\beta\mu_{1{\rm L}}=25$, $\beta\mu_{1{\rm R}}=0$, $\beta\mu_{2{\rm L}}=70$, $\beta\mu_{2{\rm R}}=0$, $\Gamma_{1{\rm L}}=\Gamma_{1{\rm R}}=\bar{\Gamma}_{1{\rm L}}= \bar{\Gamma}_{1{\rm R}} =1$, and $\Gamma_2\equiv\Gamma_{2{\rm L}}=\Gamma_{2{\rm R}}= \bar{\Gamma}_{2{\rm L}}= \bar{\Gamma}_{2{\rm R}}=1, 10, 100, 1000, 10000$.  As observed in Figs. \ref{fig3} and \ref{fig4}, the difference $Q(\lambda_1,0)-Q(\tilde{A}_1-\lambda_1,0)$ is positive for $\lambda_1<\tilde{A}_1/2$ and negative for $\lambda_1>\tilde{A}_1/2$.  Here, we only depict the difference for $\lambda_1<\tilde{A}_1/2$.  The other half has a similar structure if the absolute value of the difference is taken before the logarithm.}
\label{fig6}
\end{figure}

In the limit where the ratio of rate constants tends to infinity, the generating function becomes identical with its symmetric, as argued in Sec. \ref{Limit}.  In order to verify this prediction, we depict in Fig. \ref{fig6} the difference between both functions versus the counting parameter $\lambda_1$.  We observe in this figure that the difference is reduced by one order of magnitude each time the ratio of rate constants $\Gamma_2/\Gamma_1$ is increased by the same factor.  Consequently, the single-current fluctuation theorem is well established in the large ratio limit $\Gamma_2/\Gamma_1\to\infty$. In this limit, the effective affinity is given by Eq. (\ref{eff_A_1}).

\begin{figure}[htbp]
\centerline{\includegraphics[width=9cm]{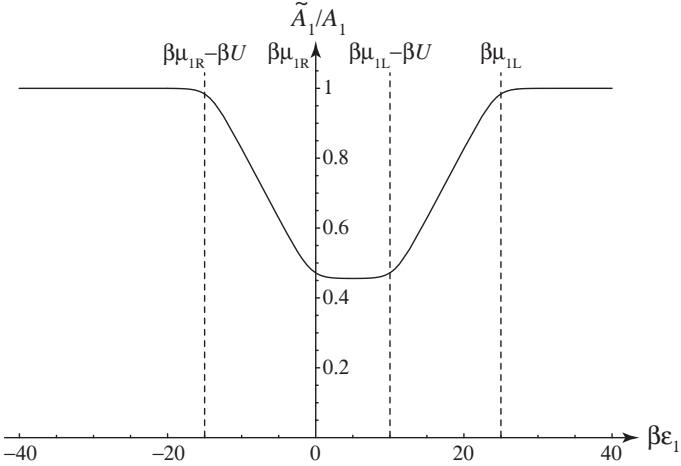}}
\caption{The effective affinity (\ref{eff_A_1}) of the QD versus the dimensionless energy $\beta\epsilon_1$ of its level for the parameter values $\beta U=15$ and (\ref{m1L})-(\ref{e2}).}
\label{fig7}
\end{figure}

The effective affinity is depicted in Fig. \ref{fig7} as a function of the energy $\beta\epsilon_1$ of the QD No.\,1 for $\beta U=15$.  We observe that the effective affinity takes the actual value (\ref{A1_act}) determined by the reservoirs for either low or large values of the energy $\beta\epsilon_1$.  However, the effective affinity undergoes a significant reduction in between, down to a minimum of about $\tilde{A}_1\simeq 0.45 \times A_1$.  The function has a characteristic shape, which can be explained in terms of the Fermi-Dirac distributions entering in the expression (\ref{eff_A_1}) of the effective affinity.  Supposing that $\mu_{\rm 2R} < \epsilon_2 < \mu_{2L}-U$ and $0<U<\mu_{\rm 1L}-\mu_{\rm 1R}$, we find that the effective affinity is approximately given by
\be
\tilde{A}_1 \simeq
\left\{
\begin{array}{ll}
\beta(\mu_{\rm 1L}-\mu_{\rm 1R}) & \mbox{for} \quad \epsilon_1<\mu_{\rm 1R}-U \\ 
\beta(-\epsilon_1-U+\mu_{\rm 1L}) & \mbox{for} \quad \mu_{\rm 1R}-U < \epsilon_1< \mu_{\rm 1R} \\ 
\beta(\mu_{\rm 1L}-\mu_{\rm 1R}-U) & \mbox{for} \quad \mu_{\rm 1R}<\epsilon_1<\mu_{\rm 1L}-U \\ 
\beta(\epsilon_1-\mu_{\rm 1R}) & \mbox{for} \quad \mu_{\rm 1L}-U < \epsilon_1< \mu_{\rm 1L} \\ 
\beta(\mu_{\rm 1L}-\mu_{\rm 1R}) & \mbox{for} \quad \mu_{\rm 1L} <\epsilon_1 \\ 
\end{array}
\right.
\label{eff_A_1_shape1}
\ee
up to corrections that are smaller than $\beta=(k_{\rm B}T)^{-1}$ in the zero temperature limit $T\to 0$.
Crossovers happen where the energy $\epsilon_1$ coincides with the values of the chemical potentials of the left- and right-hand reservoirs and the chemical potentials reduced by the Coulomb repulsion $U$.  The slope of the effective affinity versus $\beta\epsilon_1$ is successively $\{0,-1,0,+1,0\}$, as seen in Fig. \ref{fig7}.  According to Eq. (\ref{eff_A_1_shape1}), the minimum value of the effective affinity is approximately given by $\tilde{A}_1\simeq A_1-\beta U = 10$ in the middle interval $\beta\mu_{\rm 1R}=0<\beta\epsilon_1<\beta\mu_{\rm 1L}-\beta U=10$. The affinity $A_1=25$ of the reservoirs is recovered for $\beta\epsilon_1<\beta\mu_{\rm 1R}-\beta U=-15$ and for $\beta\epsilon_1>\beta \mu_{\rm 1L}=25$, which explains the features observed in Fig. \ref{fig7}.

\begin{figure}[htbp]
\centerline{\includegraphics[width=9cm]{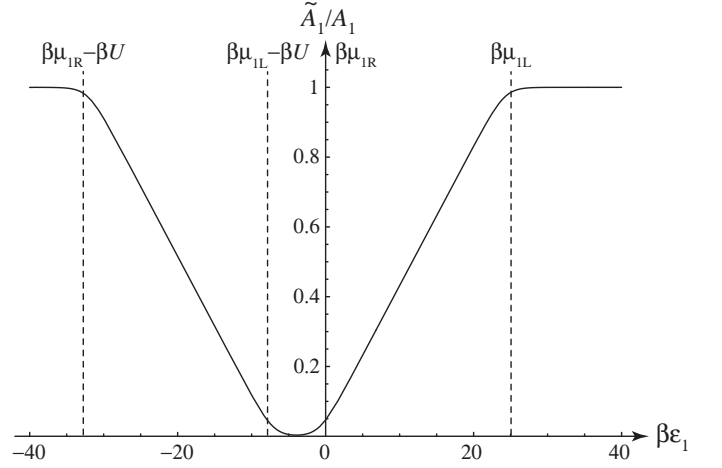}}
\caption{The effective affinity (\ref{eff_A_1}) of the QD versus the dimensionless energy $\beta\epsilon_1$ of its level for the parameter values (\ref{U})-(\ref{e2}).}
\label{fig8}
\end{figure}

Equation (\ref{eff_A_1_shape1}) predicts that the minimum value of the effective affinity could be further decreased by increasing the Coulomb repulsion $U$.  This is indeed the case as observed in Fig. \ref{fig8}, which depicts the effective affinity versus the energy $\epsilon_1$ now for the value (\ref{U}).
Here, we see that the effective affinity may vary from the maximum value given by the affinity $A_1 = 25$ imposed by the reservoirs down to the very small minimum value $\tilde A_1 \simeq 0.083565$ at $\beta\epsilon_1\simeq-3.9525$.  In particular, the value $\tilde A_1 \simeq 3$ which has been experimentally observed in Ref.~\onlinecite{FHTH06}  is reached for $\beta\epsilon_1 \simeq 2.2$.

If the condition $\mu_{\rm 2R} < \epsilon_2 < \mu_{2L}-U$ is still satisfied for the parameter values of Fig. \ref{fig8}, the Coulomb repulsion is now larger than the difference of chemical potentials:  $U>\mu_{\rm 1L}-\mu_{\rm 1R}$.  In this other regime, the effective affinity is approximately given by
\be
\tilde{A}_1 \simeq
\left\{
\begin{array}{ll}
\beta(\mu_{\rm 1L}-\mu_{\rm 1R}) & \mbox{for} \quad \epsilon_1<\mu_{\rm 1R}-U \\ 
\beta(-\epsilon_1-U+\mu_{\rm 1L}) & \mbox{for} \quad \mu_{\rm 1R}-U < \epsilon_1< \mu_{\rm 1L}-U \\ 
0 & \mbox{for} \quad \mu_{\rm 1L}-U<\epsilon_1<\mu_{\rm 1R} \\ 
\beta(\epsilon_1-\mu_{\rm 1R}) & \mbox{for} \quad \mu_{\rm 1R} < \epsilon_1< \mu_{\rm 1L} \\ 
\beta(\mu_{\rm 1L}-\mu_{\rm 1R}) & \mbox{for} \quad \mu_{\rm 1L} <\epsilon_1 \\ 
\end{array}
\right.
\label{eff_A_1_shape2}
\ee
up to corrections that are smaller than $\beta=(k_{\rm B}T)^{-1}$ in the zero temperature limit $T\to 0$.
In the middle interval $\beta\mu_{\rm 1L}-\beta U=-7.8<\beta\epsilon_1<\beta\mu_{\rm 1R}=0$, the minimum effective affinity reaches a value that vanishes in the low temperature limit $T\to 0$.  The actual value of the affinity $A_1=25$ is recovered for $\beta\epsilon_1<\beta\mu_{\rm 1R}-\beta U=-32.8$ or $\beta\epsilon_1>\beta \mu_{\rm 1L}=25$.

\begin{figure}[htbp]
\centerline{\includegraphics[width=9cm]{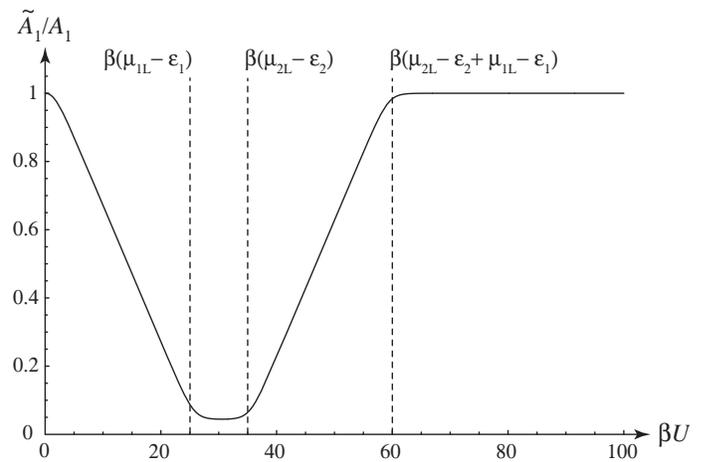}}
\caption{The effective affinity (\ref{eff_A_1}) of the QD versus the dimensionless electrostatic coupling constant $\beta U$ for the parameter values $\beta\epsilon_1=0$ and (\ref{m1L})-(\ref{e2}).}
\label{fig9}
\end{figure}

The dependence of the effective affinity (\ref{eff_A_1}) on the Coulomb repulsion is shown in Fig. \ref{fig9} for a given value of the energy $\beta\epsilon_1=0$.  Here also, the effective affinity can be reduced down to a much lower value than the one determined by the reservoirs.  By a reasoning similar to the one used to get Eqs. (\ref{eff_A_1_shape1}) and (\ref{eff_A_1_shape2}), we can obtain the approximate dependence of the effective affinity on the parameter $U$ under the conditions $\mu_{\rm 1R}-\epsilon_1 <0<\mu_{\rm 1L}-\epsilon_1 < \mu_{\rm 2L}-\epsilon_2$ as follows:
\be
\tilde{A}_1 \simeq
\left\{
\begin{array}{l}
\beta(-U+\mu_{\rm 1L}-\mu_{\rm 1R}) \\ \quad\qquad\mbox{for} \quad 0 < U < \mu_{\rm 1L}-\epsilon_1 \\ 
\beta(\epsilon_1-\mu_{\rm 1R}) \\ \quad\qquad\mbox{for} \quad \mu_{\rm 1L}-\epsilon_1 < U < \mu_{\rm 2L}-\epsilon_2 \\ 
\beta(U+\epsilon_2-\mu_{\rm 2L}+\epsilon_1-\mu_{\rm 1R}) \\ \quad\qquad\mbox{for} \quad \mu_{\rm 2L}-\epsilon_2 < U < \mu_{\rm 2L}-\epsilon_2+\mu_{\rm 1L}-\epsilon_1 \\ 
\beta(\mu_{\rm 1L}-\mu_{\rm 1R}) \\ \quad\qquad\mbox{for} \quad \mu_{\rm 2L}-\epsilon_2+\mu_{\rm 1L}-\epsilon_1 < U \\ 
\end{array}
\right.
\label{eff_A_1_shapeU}
\ee
up to corrections smaller than $\beta=(k_{\rm B}T)^{-1}$ as $T\to 0$.
The piecewise linear approximation obtained from the Fermi-Dirac distributions here also explains the successive slopes $-1$, $0$, $+1$, and $0$, observed in the plot of the effective affinity versus $\beta U$.
We notice that the different linear pieces of the approximation match together at the crossover values of the variable $\beta U$.  The minimum value is reached in the interval $\beta(\mu_{\rm 1L}-\epsilon_1)=25 < \beta U < \beta(\mu_{\rm 2L}-\epsilon_2)=35$ while the affinity $A_1=25$ of the reservoirs is recovered for $\beta U > 
\beta(\mu_{\rm 2L}-\epsilon_2+\mu_{\rm 1L}-\epsilon_1)=60$, as indeed confirmed by Fig. \ref{fig9}.

The lowering of the effective affinity under specific conditions can be explained in the present model as the effect of the back action of the secondary circuit interacting with the observed quantum dot.  The charging and discharging rates of the quantum dot can be drastically modified by the coupling to the secondary circuit.  In this way, the effective affinity can be much reduced in some regimes which are determined by the value of the energy $\epsilon_1$ of the quantum dot with respect to the values of the chemical potentials and the electrostatic coupling parameter $U$. This back-action effect tends to disappear as the temperature increases at constant voltages.  

We emphasize that the reduction of the effective affinity is not caused by the Coulomb drag.  Indeed, there is no Coulomb drag for the conditions chosen in the present section because we have here taken rate constants such that $\Gamma_j=\bar{\Gamma}_j$ as in Eqs. (\ref{G1}) and (\ref{G2}).  Therefore, the Onsager coefficient (\ref{L12_prop}) vanishes together with higher-order coefficients according to Eqs. (\ref{J_a=0}) and (\ref{LMN}) and the Coulomb drag does not manifest itself for the conditions taken in Figs. \ref{fig3}-\ref{fig9}.

\section{Conclusions}
\label{Conclusions}

In the present paper, we have reported the study of the single-current fluctuation theorem in a Hamiltonian model of quantum electron transport in two capacitively coupled channels, each containing a quantum dot (QD).\cite{SKB10}  Such a system is similar to the electronic devices used in typical counting statistics experiments\cite{FHTH06,GLSSISEDG06} where the current in one circuit can continuously monitor the state of the QD in the other circuit thanks to the capacitive coupling.  The model allows us to investigate the effects of the back action of the monitoring circuit on the counting statistics in the light of the so-called fluctuation theorems.

Since both circuits are capacitively coupled and microreversibility holds for the total Hamiltonian (\ref{H}), a fluctuation theorem is satisfied for the two currents flowing across the system.  This two-current fluctuation theorem (\ref{FT}) or (\ref{FT-p}) relates the counting statistics of opposite random electron transfers in both circuits to the affinities or thermodynamic forces (\ref{A1})-(\ref{A2}) driving the system away from equilibrium.
The fluctuation theorem is valid far from equilibrium in the strongly nonlinear regimes encountered in electronic circuits composed of quantum dots and quantum point contacts.

However, in counting statistics experiments, one circuit is used to monitor the current fluctuations in the other circuit so that the counting statistics cannot be carried out on both currents together and is thus restricted to a single current.  Accordingly, such experiments can only test a single-current fluctuation theorem.
In general, the two-current fluctuation theorem does not imply the single-current fluctuation theorem except under certain conditions\cite{AG07} or in some limits as we have demonstrated in the present paper.

In Sec. \ref{Uinfty}, we have studied the limit of large capacitive coupling between both circuits.  In this limit, the state of simultaneous occupancy of both QDs in the two parallel channels is at a so high energy that it is energetically forbidden.  The consequence is that the two single-occupancy states are separately accessible only from the empty state and the single-current fluctuation theorem holds with respect to the affinity determined by the electrochemical potentials of the reservoirs.

In Sec. \ref{Limit}, we have instead considered the limit where the current in one circuit is much larger than in the other circuit.  Indeed, a large current ratio is a key feature of typical counting statistics experiments\cite{FHTH06,GLSSISEDG06} where the current ratio reaches values as high as $10^7$-$10^8$.  The circuit with the very large current performs the continuous-time monitoring of the quantum state of the QD in the other circuit.  In such a limit, the charging and discharging rates of the slow QD take values averaged over the very fast fluctuations of the monitoring circuit.  This is the essence of the back action of the monitoring circuit onto the QD circuit.  As a consequence of the large current ratio limit, the single-current fluctuation theorem holds but with respect to the effective affinity (\ref{eff_A_1}), which can significantly differ from the actual value of the affinity determined by the reservoirs of the corresponding circuit.  This modification of the affinity is due to the capacitive coupling between both circuits, as shown in particular by Eq. (\ref{eff_A_1_shape1}).  In terms of the parameter $U$ of the Coulomb electrostatic interaction appearing in the Hamiltonian (\ref{H_S}), the affinity is lowered according to $\tilde{A}_1 \simeq A_1 - \beta U$ under the conditions specified around Eq. (\ref{eff_A_1_shape1}).  This result explicitly expresses the effect of the back action between both circuits on the single-current fluctuation theorem.  This back-action effect can be reduced if the Coulomb repulsion $U$ is decreased, but the monitoring circuit can no longer resolve the two states of the QD as in Fig. \ref{fig2} if $U$ is too small.  On the other hand, the back-action effect is also reduced for large values of the Coulomb repulsion as shown by Eq. (\ref{eff_A_1_shapeU}) and in Fig. \ref{fig9}.  Indeed, for a large Coulomb repulsion, the affinity recovers the value determined by the reservoirs and the back-action effect disappears.  This case corresponds to the situation considered in Ref.~\onlinecite{CTH10} where a quantum fluctuation theorem has been obtained in a multiple measurements scheme.

From a general viewpoint, the two-current fluctuation theorem implies the non-negativity of the entropy production in agreement with the second law of thermodynamics. The dissipation of energy can thus be evaluated in the electron transport process used to perform quantum measurement in the experiments of Refs.~\onlinecite{FHTH06,GLSSISEDG06}.  This dissipation of energy accompanying quantum measurement is expected on fundamental ground.\cite{vN55}  The necessity of resolving the QD state in real time has for direct consequence that the dissipation in the monitoring circuit is much higher than in the QD by a factor $\Pi_2/\Pi_1=(A_2/A_1)\times(J_2/J_1)$ of the same order of magnitude as the current ratio $J_2/J_1$. If the QD state is monitored with a sampling time $\Delta t$, the secondary circuit playing the role of the detector should have transitions on equal or shorter time scales according to Eq. (\ref{Dt_real}).  Since the secondary circuit is driven out of equilibrium by the affinity $A_2$, its electron current should satisfy $J_2 \gtrsim (\Delta t)^{-1}$, so that the dissipated power should be bounded by $\Pi_2=k_{\rm B}T A_2J_2 \gtrsim k_{\rm B}T A_2(\Delta t)^{-1}$.  The higher the time resolution, the higher the dissipation rate.

In summary, we have shown that the single-current fluctuation theorem is valid under different limiting conditions and provided a fundamental understanding of the back-action effect of the monitoring circuit on the affinity of the monitored circuit, as observed in Ref.~\onlinecite{FHTH06}.  The present study extends the analysis of Ref.~\onlinecite{UGMSFS10,UGMSFS10bis} in showing how the effective affinity of the single-current fluctuation theorem can be directly expressed in terms of the parameters entering the Hamiltonian of the system.  Several issues are left open such as the facts that a quantum point contact has specific transport properties and that the counting statistics is performed by two QDs in series in the experiments of Ref.~\onlinecite{FHTH06}.  We hope to report on these issues in a forthcoming publication.

\begin{acknowledgments}
The authors are grateful to David Andrieux for fruitful discussions as well as for communicating them his results of July 2008 on the estimation of the effective affinity for the experiment reported in Ref.~\onlinecite{FHTH06}.
G. Bulnes Cuetara thanks the ``Fonds pour la Formation \`a la Recherche dans l'Industrie et l'Agriculture" (FRIA Belgium) for financial support.
M. Esposito is supported by the Belgian Federal Government under the Interuniversity Attraction Pole project ``NOSY" and by the European Union Seventh Framework Programme (FP7/2007-2013) under grant agreement 256251.
\end{acknowledgments}

%%%%%%%%%%%%%%%%%%%%%%%%%%%%%%%%%%%%%%%%%%%%%%%%%%
\appendix

\section{Calculation of the average currents}
\label{AppA}

In this appendix, two different methods are developed in order to calculate the average currents given by Eq. (\ref{J_a}) in terms of the leading eigenvalue $Q$ of the eigenvalue problem (\ref{eigenvalue_problem}) of the four-by-four matrix (\ref{L}).  

The first method starts from the eigenvalue equation (\ref{eigenvalue_problem}) for the right eigenvector $\bf v$ associated with the eigenvalue $Q$ and from the adjoint equation
\be
\mbox{\helvb L}^{\rm T} \cdot {\bf u} = - Q  \, {\bf u} 
\label{left}
\ee
for the left eigenvector $\bf u$ where $^{\rm T}$ denotes the transpose of the matrix.  The left and right eigenvectors satisfy the normalization condition
\be
{\bf u}^{\rm T}\cdot{\bf v}=1
\label{u-v}
\ee
Accordingly, the eigenvalues is given by
\be
Q = - {\bf u}^{\rm T}\cdot\mbox{\helvb L}\cdot{\bf v}
\label{Q-u-v}
\ee
Taking the partial derivative $\partial_\alpha$ with respect to the counting parameter $\lambda_{\alpha}$ of Eqs. (\ref{u-v})-(\ref{Q-u-v}) and using Eq. (\ref{eigenvalue_problem}) and Eq. (\ref{left}), we obtain the following expression for the average current:
\be
J_\alpha =\left. \partial_\alpha Q\right\vert_{\pmb{\lambda}=0} =\left. - {\bf u}^{\rm T}\cdot\partial_\alpha\mbox{\helvb L}\cdot{\bf v}\right\vert_{\pmb{\lambda}=0}
\ee 
Since the left and right eigenvectors are given at $\pmb{\lambda}=0$ by
\be
\left.{\bf u}\right\vert_{\pmb{\lambda}=0}=
\left(
\begin{array}{c}
1 \\
1 \\
1 \\
1
\end{array}
\right)
\qquad\mbox{and}\qquad
\left.{\bf v}\right\vert_{\pmb{\lambda}=0}=
\left(
\begin{array}{c}
P_{00} \\
P_{10} \\
P_{01} \\
P_{11}
\end{array}
\right)
\ee
in terms of the probabilities (\ref{Probs}), we get Eqs. (\ref{J1}) and (\ref{J2}) for the average currents. 

With the second method, the average currents as well as the linear response coefficients are directly calculated in terms of the characteristic determinant (\ref{det}) of the four-by-four matrix (\ref{L}).  This determinant is a polynomial of fourth degree:
\be
Q^4 + C_3 Q^3 + C_2 Q^2 + C_1 Q + C_0 = 0
\ee
where the coefficients depend on the parameters of the model as well as on the counting parameters $\lambda_1$ and $\lambda_2$. We notice that the last coefficient is just the determinant of the matrix (\ref{L}): $C_0=\det\mbox{\helvb L}$.

Since the matrix (\ref{L}) reduces to the matrix of a jump stochastic process conserving probability if $\lambda_1=\lambda_2=0$, the leading eigenvalue vanishes in this limit:
\be
Q(0,0)=0
\label{Q=0}
\ee
Since the average currents are given by Eq. (\ref{J_a}), we take the partial derivative $\partial_\alpha$ of the characteristic determinant with respect to the counting parameter $\lambda_\alpha$ to get 
\bea
&&\left(4Q^3 + 3C_3 Q^2 + 2C_2 Q + C_1\right)\partial_\alpha Q \nonumber\\
&&+\, \partial_\alpha C_3 \, Q^3 + \partial_\alpha C_2 \, Q^2 + \partial_\alpha C_1 \, Q + \partial_\alpha C_0=0 \nonumber\\
\label{D_aQ}
\eea
Now, the counting parameters must be set equal to zero and, according to Eq. (\ref{Q=0}), the average current is thus given by
\be
J_\alpha = - \left.\frac{\partial_\alpha C_0}{C_1}\right\vert_{\pmb{\lambda}=0}
\ee

Using the symbolic manipulation software Mathematica,\cite{Mathematica} we can evaluate the derivative $\partial_\alpha C_0$ if the reservoirs of the circuit $\alpha$ are at equilibrium, i.e., if its electrochemical potentials are equal so that its affinity is vanishing: $A_\alpha=\beta(\mu_{\alpha{\rm L}}-\mu_{\alpha{\rm R}})=0$.  The result is that this quantity vanishes under the condition $\Gamma_j = \bar{\Gamma}_j$ even if the other circuit is out of equilibrium, which establishes Eq. (\ref{J_a=0}).

We notice that the Onsager coefficient can also be obtained in the same way.  Using Eq. (\ref{L12}) and taking a further derivative of Eq. (\ref{D_aQ}) with respect to the other counting parameter $\lambda_\beta$, we get
\be
L_{\alpha,\beta}=L_{\beta,\alpha}= \left.\frac{\partial_\alpha \partial_\beta C_0}{2C_1}\right\vert_{\pmb{\lambda}=0,{\bf A}=0}
\label{L12_det}
\ee
which is used to obtain Eq. (\ref{L12_prop}) with the symbolic manipulation software Mathematica.\cite{Mathematica} 

%%%%%%%%%%%%%%%%%%%%%%%%%%%%%%%%%%%%%%%%%%%%%%%%%%

\end{document}